\documentclass[annual]{acmsiggraph}

\TOGonlineid{0268}

\graphicspath{{./figures/}}

\title{Mesh2Fab: Reforming Shapes for Material-specific Fabrication}

\author{Yong-Liang Yang$^{1,2}$\hspace{.4in}
        Jun Wang$^1$\hspace{.4in}
        Niloy J. Mitra$^3$\\
        $^1$KAUST\hspace{.4in}
        $^2$University of Bath\hspace{.4in}
        $^3$UCL}

\pdfauthor{}


\let\vec=\mathbf

\let\set=\mathcal

\newcommand{\mypara}[1]{\paragraph{#1.}}

\newcommand{\vnudge}{\vspace{-.1in}}

\def \path {\mathit{path}}

\def \label {\mathit{label}}

\let\rm=\rmfamily    \let\sf=\sffamily    
     \let\sl=\slshape     \let\sc=\scshape
\let\bf=\bfseries

\newcommand{\denselist}{\itemsep 0pt\parsep=1pt\partopsep 0pt}
\newcommand{\bitem}{\begin{itemize}\denselist}
\newcommand{\eitem}{\end{itemize}}
\newcommand{\benum}{\begin{enumerate}\denselist}
\newcommand{\eenum}{\end{enumerate}}
\newcommand{\bdescr}{\begin{description}\denselist}
\newcommand{\edescr}{\end{description}}

\usepackage{parskip}

\usepackage{algorithm2e}
\usepackage{wrapfig}

\begin{document}

 \teaser{
   \includegraphics[width=.95\textwidth]{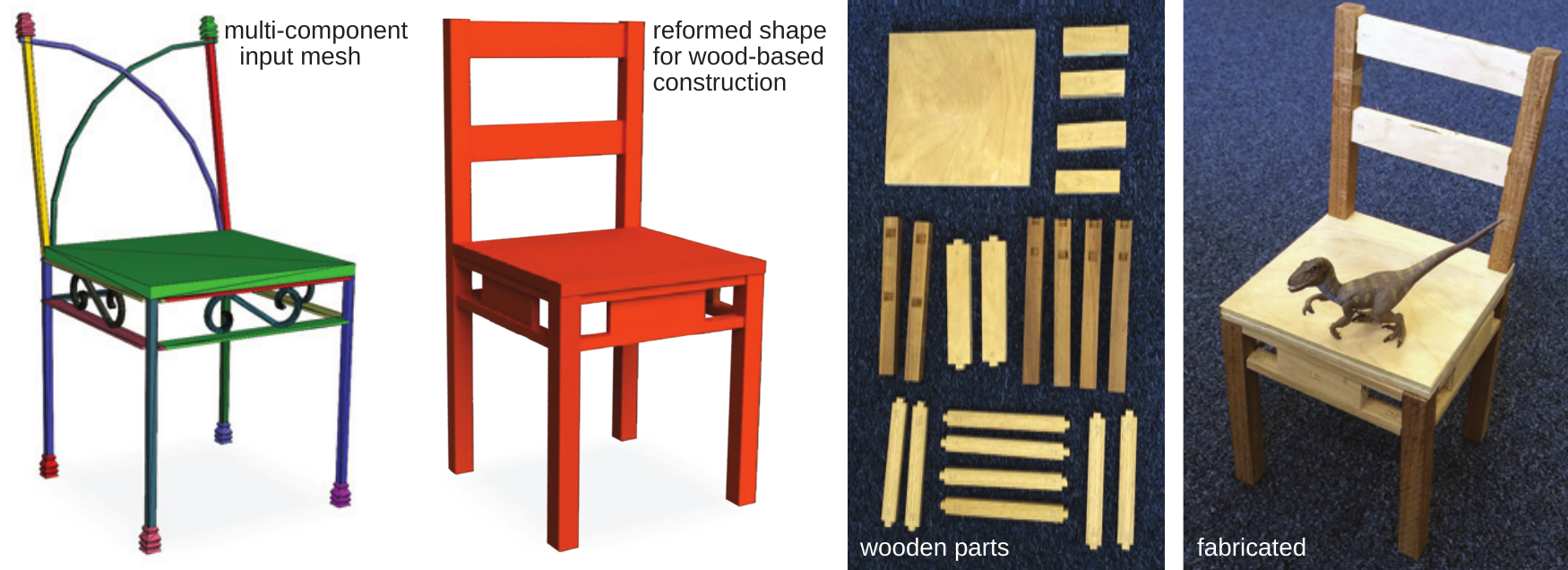}
   \caption{We present Mesh2Fab, a framework to reform (i.e, reshape) an input multi-component mesh to simplify fabrication from a target built material. In this example, the input model~(left) is modified both topologically and geometrically leading to a reformed shape~(middle-left) along with necessary part and joint information (middle-right), which can then be used to fabricate the shape~(right). }
   \label{fig:teaser}
   \vnudge
 }

\maketitle

\begin{abstract}
As humans, we regularly associate shape of an object with its built material.
In the context of geometric modeling, however, this inter-relation between form and material is rarely explored.
In this work, we propose a novel data-driven reforming (i.e., reshaping) algorithm that adapts an input multi-component model for a target fabrication material.
The algorithm adapts both the part geometry and the inter-part topology of the input shape to better align with material-specific fabrication requirements. 
 As output, we produce the reshaped model along with respective part dimensions and inter-part junction specifications. 
We evaluate our algorithm on a range of man-made models and demonstrate non-trivial model reshaping examples focusing only on metal and wooden materials. We also appraise the output of our algorithm using a user study.
\end{abstract}

\keywordlist



\section{Introduction}
\label{sec:intro}

\begin{figure}[b!]
\vnudge
  \centering
  \includegraphics[width=\columnwidth]{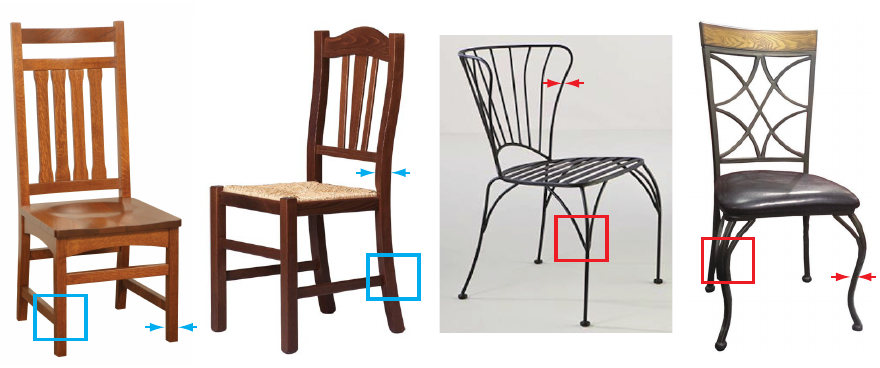}
  \caption{Geometric properties such as thickness of individual parts and contact angles between parts are characteristics of corresponding built materials. Blue box highlights near right angles between wooden parts, while red box highlights more flexible angles between metal parts.}
  \label{fig:fabrication_angle}
\end{figure}

Geometric form of a physical object is strongly dictated by its built material.
This is not surprising since materials differ as to how they can be warped towards a geometric form.
For example, it is desirable, both in terms of increased convenience and reduced waste, to cut wood into straight planks, while
others like metal sheets or plywood can easily be given a simple curved profile without much additional cost.
Thus built materials affect both the shape of the parts and how they can be interconnected.
As humans, we often correlate geometric appearance of an object, even in absence of any texture or color information,  with its fabrication material (see Fig.~\ref{fig:fabrication_angle}).

In geometric modeling, shape and physical material are rarely considered together. Traditionally, as graphics objects are largely used in virtual environments, such an approach is entirely justified.
Moreover, one can attach any (virtual) material texture to any geometric shape.
Only recently, with growing interest in actual fabrication to bring virtual objects back to the real world, fabrication-aware modeling has gained popularity~\cite{Bic10,Li:2010:PAP,HildebrandBA12,Prevost:2013:MSB,Schwartzburg:2013:FDI}.
Such methods, however, focus on rationalization (i.e., close approximation) of a target shape by deriving an economic and feasible solution. 
In contrast, we are interested in reforming shapes to simplify subsequent fabrication.

Given an input shape, we are interested in how it will look if fabricated using a target built material. Essentially, we investigate the implicit relation
between shape and built material (c.f., {\em material cause} of Aristotle's concept of causes).
For example, the model shown in Fig.~\ref{fig:teaser}-left, which has thin parts connected at narrow angles, would be difficult to fabricate in its original shape with wood. A reformed shape with thicker parts meeting together at near-orthogonal angles is much more suited for wood-based fabrication (Fig.~\ref{fig:teaser}-middle).
We present a data-driven reshaping algorithm for this purpose that suggests a new shape along with necessary fabrication specifications (Fig.~\ref{fig:teaser}-right).

\begin{figure*}[t!]
  \centering
  \includegraphics[width=.95\linewidth]{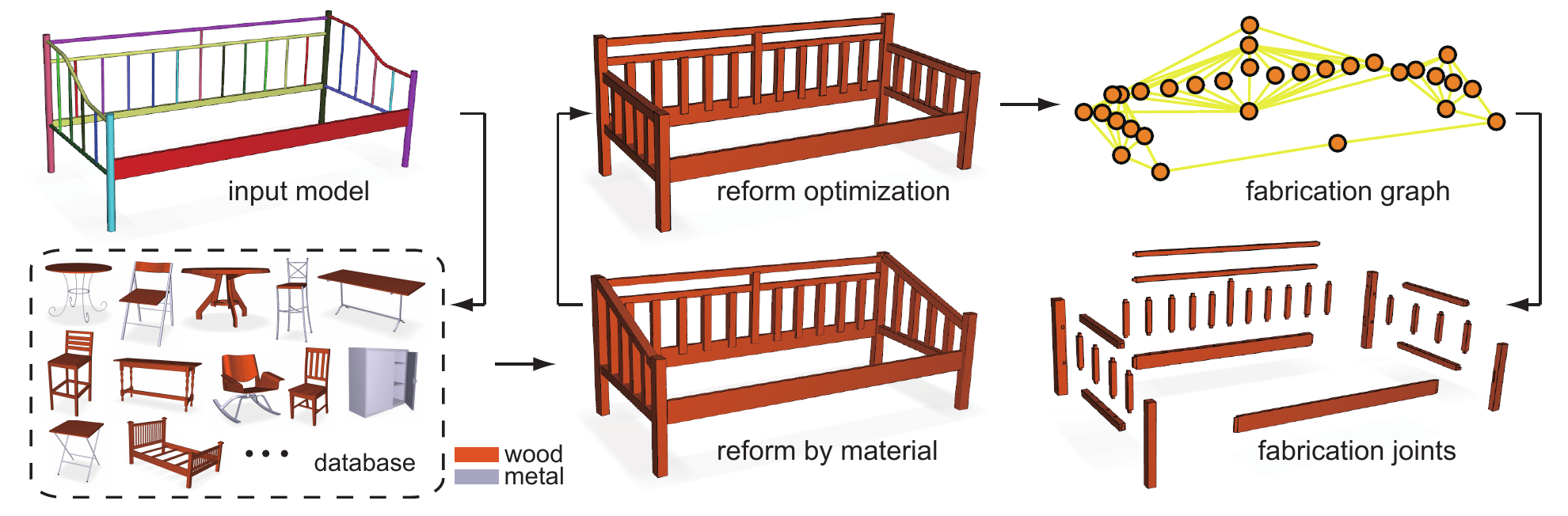}
  \caption{Overview of our framework.}
  \label{fig:overview}
\end{figure*}

Starting from a set of objects with part materials assigned, in a preprocessing stage, we extract the correlation between shape of parts and their assigned raw material in the context of component-based man-made objects. Note that we do {\em not} assume the components to have semantic labels (chair leg, table top, etc.).
In this paper, we only consider two types of materials: wood and metal.
Based on the learned information, we adapt input models to suit fabrication from a target material.
Since different materials favor different part pair contact profiles (i.e., contact angles), the proposed algorithm adapts the parts to better conform to preferred contact profiles.
This involves both topological adaptations by breaking old and creating new contacts, and geometric adaptations by solving for appropriate new contact locations for corresponding part profiles.
The algorithm outputs both part dimensions and inter-part connection specifications to simplify physical fabrication of the reshaped objects.
We evaluate our framework on a range of data sets and demonstrate its efficacy for novel material-driven object reshaping. The created forms are also validated by a user study.

In summary, our contributions include:
\begin{itemize}
  \item formulating a computational connection between geometric form and physical fabrication material; and
  \item introducing a novel algorithm to reshape input objects towards specified built material.
\end{itemize}

\section{Related Work}
\label{sec:relatedWk}

\mypara{Fabrication-aware modeling}
Rapid advances in accessible and economic fabrication possibilities have renewed interest in
fabrication-aware modeling. Starting from an input shape and construction material,
the goal of such methods is to best approximate the input shape while conforming to constraints arising due to   fabrication methodology.
Examples include
incorporating curved folds to create freeform surfaces for folding
single planar sheet of material~\cite{kfcmsp_folding_sig_08};
rationalization of freeform surfaces with triangular, quad, cylindrical, and other primitive panels~\cite{Eigensatz:2010:PAF,Fu:2010:KTS,Singh:2010:TSD};
designing cardboard chairs with stability considerations~\cite{Saul:2010:SAC};
fabricate materials with target behavior~\cite{Bic10};
generating fabricatable parts and connectors from an input wooden furniture model using a grammar-based method~\cite{Lau:2011:CFM}; or adapting input models to facilitate construction from planar pieces~\cite{HildebrandBA12,Schwartzburg:2013:FDI}.
%
 We aim at reforming shapes for material-specific fabrication. Our method can adapt a multi-component model to a different fabrication context (e.g., metal$\rightarrow$wood, wood$\rightarrow$metal) using pre-knowledge from a database. To the best of our knowledge, we are not aware of prior work investigating this problem.

\mypara{Shape deformation and synthesis}
Shape deformation deals with warping an input shape based on user provided positional specifications
while trying to best maintain certain model properties. Such properties can preserve
local geometric details (c.f.,~\cite{Botsch:2008:LVS}); regularize deformations to be as-rigid-as-possible~\cite{Sorkine:2007:ASM} or near-isometric~\cite{kmp_shape_space_sig_07};
or conform to inter- and intra-part relations analyzed from the input models~\cite{gsmc_iwires_sig_09}.
In the context of content creation, there exists different methods to synthesize model variations
starting from a collection of input shapes. For example, modeling can be performed by
mixing-and-matching among model parts~\cite{Funkhouser:2004:MBE,shuffler_07}; sampling from a learned part-based probability distribution~\cite{Chaudhuri:2011:PRA,Kalogerakis:2012:PMC};
using an evolutionary algorithm to create model variations~\cite{Xu:2012:FDS};
or exchanging compatible part substructures to create plausible model variations~\cite{Zheng:2013:FSP}.
More recently, algorithms have been proposed to minimally change input shapes such that the fabricated
final shape is  resilient and physically stable~\cite{Stava:2012:SRI,Umetani:2012:GEP,Prevost:2013:MSB}.
In contrast, we explore the relation between material and form, and how they influence each other resulting in large form changes, both topological and geometric.


\section{Overview}
\label{sec:overview}
Given a query model and user-specified target materials for individual parts, our goal is to use the query model as reference, and generate a new model subject to prescribed material constraints (see Fig.~\ref{fig:teaser}). The main idea is to synthesize new geometric forms based on example models in a database. First, the new model parts are extracted from the database under the guidance of the input model. However, due to material change, the new part configurations (e.g., the contact angle enclosed by neighboring parts) need to be revised in an optimization step, to facilitate fabrication in the target material context. Moreover, fabrication specifications (e.g., woodworking joints) can also be inferred from the database, making the real fabrication feasible. The whole pipeline of our framework is illustrated in Fig.~\ref{fig:overview}.


\section{Algorithm}


\subsection{Preprocessing}
\mypara{Data preparation}
The input to our framework is a multi-component model $\mathcal{P}$ expressed as a set of parts $\{P_1, P_2, ... , P_N\}$.
Note that we assume only part decomposition, but make no assumption about part labeling or correspondence in the input.
Each part is represented by a triangular mesh. If there is no pre-knowledge, a part segmentation based on triangle face connectivity is performed. The user may further divide and/or re-group parts if necessary. Further, there is a database which contains 152 multi-component models from Google 3D Warehouse.
For the models in the database, we also tag a fabrication material (wood/metal/other) for each part. The parts tagged with `other' are ignored in the analysis, so as small parts which are not key elements of the model (e.g. screw, chair leg pad, etc.).

\mypara{Part analysis}
For each model part, we compute an oriented bounding box (OBB) using principle component analysis of all mesh vertices. To overcome irregular triangulation, we further optimize the OBB by iteratively fixing one dimension and optimizing the other two dimensions using rotating calipers~\cite{Toussaint:1983:RC}. Uniform scaling is applied to each model according to the diagonal length of its OBB. We also find part thickness is closely related to fabrication material, and perform thickness estimation as follows. We first sample points (1000 in our experiments) on mesh surface. Starting from each sample point $\vec{s}_i$, we shoot a ray along its normal (i.e., the normal of the mesh face where the point resides), and find the first intersection point $\vec{t}_i$ that hits the part. The distance $d_i=\|\vec{s}_i-\vec{t}_i\|$ is defined as the thickness of sample point $\vec{s}_i$. We then perform a voting over all sample points. The distance value $d^{\star}$ which gets the most votes is defined as the part thickness $t$.

\begin{figure} [h]
  \centering
  \includegraphics[width=.8\columnwidth]{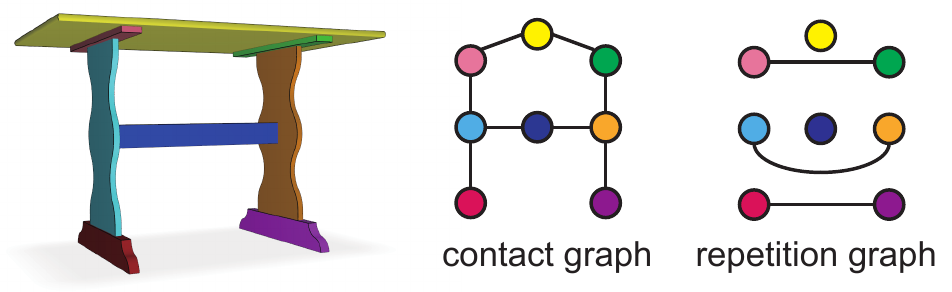}
  \caption{A simple table model along with its contact graph and repetition graph.}
  \vnudge
  \label{fig:preprocess}
\end{figure}

\mypara{Contact analysis}
For a given model, we build a contact graph $G_c=(\set{V}, \set{E}_c)$ to encode the spatial relation among parts (see Fig.~\ref{fig:preprocess}). Each node $v_i \in \set{V}$ represents a single part. If the minimum distance between two parts, $P_i$ and $P_j$, is less than a given threshold $d_c$,  we add an edge $e_{ij} \in \set{E}_c$ connecting the corresponding nodes in $G_c$. For two parts that are in contact, we also store a contact point $\vec{c}_{ij}$. We densely sample the two contact parts and extract nearby points (within $d_c$). The contact point is estimated by the barycenter of the nearby points.

\mypara{Repetition detection}
 We also build a repetition graph $G_r$ to encode the part repetitions (see Fig.~\ref{fig:preprocess}). We detect congruent parts by aligning their OBB's and measuring root mean square distance. The repetition graph $G_r=(\set{V}, \set{E}_r)$ has the same nodes as $G_c$, while each edge $e_{ij} \in \set{E}_r$ connects two congruent parts.

\subsection{Evaluating Similarity}
\label{subsec:similarity_metrics}
To reform to the new model with help of the database, we correlate the query model with the example models using a set of similarity metrics. These metrics capture the similarity in terms of part geometry, part material, and the configuration of neighboring parts, which help to adapt to the new materials while respecting the original structure of the model.


%

\mypara{Shape similarity}
We define shape similarity $\rho_{shp}(P_i, P_j)$ of two parts $P_i$ and $P_j$ based on part's OBB, area, and thickness. First, we measure the bounding box similarity as $\rho_{obb}(P_i, P_j)=\exp(-|\vec{b}_i-\vec{b}_j|^2 / \sigma_{\vec{b}}^2),$
 where $\vec{b}_i$ is a vector with 3 entries that represents $P_i$'s size from its OBB. We sort the 3 entries and use $L1$ norm to measure the difference. Second, suppose $A_P$ is the area of part $P$ and $A_{obb}$ is the area of its OBB, we also compute an area ratio $r=A_P / A_{obb}$ of each part. The area ratio similarity is defined as $\rho_{area}(P_i, P_j)=\exp(-(r_i-r_j)^2 / \sigma_r^2)$. We also define a thickness similarity as $\rho_{thick}(P_i, P_j)=\exp(-(t_i-t_j)^2 / \sigma_t^2)$. Note that sophisticated geometric descriptors (e.g., shape distribution~\cite{Osada:2002:SD}) can also be used as alternative, but may restrict suitable form changes when adapting to different materials.

%



\mypara{Material similarity}
The material similarity between $M_i$ and $M_j$ ($M_i, M_j \in \{\text{wood, metal}\}$) is simply defined as: $$\rho_{mat}(M_i, M_j) = \begin{cases} 0, & \mbox{if } M_i \neq M_j \\ 1, & \mbox{if } M_i=M_j. \end{cases}$$

\mypara{Spatial similarity}
Besides part similarity, we also define similarity for part pairs to correlate forms in a structural context. First, we consider the spatial relation between a pair of parts $P_i$ and $P_j$. We use the Euclidean distance $d_{i,j}$ between part barycenters because it is independent of model orientations. The spatial similarity between two pairs of parts is defined as:
$$\rho_{pr}(d_{i,j}, d_{k,l})=\exp({-(d_{i,j}-d_{k,l})^2}/{\sigma_{pr}^2}).$$

\mypara{Contact angle similarity}
How to connect two parts and in what form they should contact largely depend on their fabrication material (see Fig.~\ref{fig:fabrication_angle}). An important observation is made for the contact angles between linear/curvilinear parts. Wooden parts usually contact in right angles (i.e., 90 degrees), while metal parts can form flexible contact angles. This can help us to better understand the correlation between material and form. To estimate contact angle, we first identify whether the contact part is linear/curvilinear based on its OBB. If the OBB is elongated and the area ratio is near to 1, we use the maximal dimension as one side of the angle. Otherwise if the area ratio is small, we collect the sample points (those for detecting the contact point) in a local neighborhood centered at the contact point, and compute dominant principal direction to enclose the angle.
If one of the two parts is not linear/curvilinear, we mark the contact angle as N/A. For simplicity, we do not separate acute and obtuse angles, i.e., $\alpha_{ij} \in [0, 90]^{\circ}$. The contact angle similarity is defined as:
$$
\rho_{ca}(\alpha_{i,j}, \alpha_{k,l}) =
     \begin{cases}
       1 & \alpha_{i,j}, \alpha_{k,l} \text{ N/A}\\
       0 & \alpha_{i,j} \in [0, 90]^{\circ}, \alpha_{k,l} \text{ N/A}\\
       0 & \alpha_{i,j} \text{ N/A}, \alpha_{k,l} \in [0, 90]^{\circ}\\
       \exp(\frac{-(\alpha_{i,j}-\alpha_{k,l})^2}{\sigma_{ca}^2}) &\alpha_{i,j}, \alpha_{k,l} \in [0, 90]^{\circ}. \
     \end{cases}
$$

 In our experiments, plausible results are achieved with $\sigma_{\vec{b}}=0.1, \sigma_r=0.1, \sigma_t=0.02, \sigma_{pr}=0.2, \sigma_{ca}=10$. More advanced weight learning strategy~\cite{Lin:2013:PCS} can be used to further improve the results.

\subsection{Reforming Shapes based on Target Material}
Given a model $\mathcal{P}$ consists of individual parts $\{P_1, P_2, ... , P_N\}$, the goal of shape reform is to adapt the geometry of the model when changing materials to $\{\bar{M}_1, \bar{M}_2, ..., \bar{M}_N\}$. The key idea is to use similar parts from the pre-tagged database $\{P_1^t, P_2^t, ..., P_K^t\}$ (with materials $\{M_1^t, M_2^t, ..., M_K^t\}$) to replace the model parts subject to material constraints. We formulate the shape reform problem as an optimization, which jointly maximizes part similarity and pairwise similarity referring to the database.

\begin{figure}[t!]
  \centering
  \includegraphics[width=.95\columnwidth]{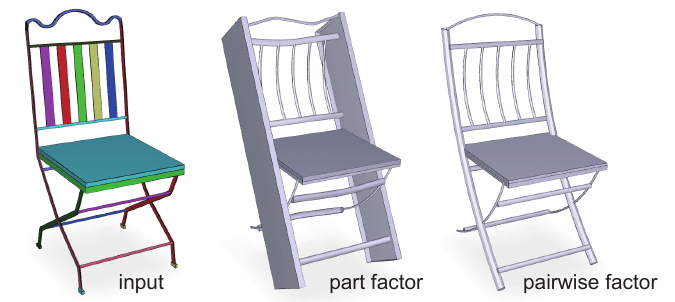}
  \caption{Pairwise factor in the shape reform optimization is based on comparing neighboring configuration in a database. This helps to suggest right geometric configuration between parts.}
  \label{fig:reform_comparison}
\end{figure}

\begin{figure}[b!]
  \centering
  \includegraphics[width=\columnwidth]{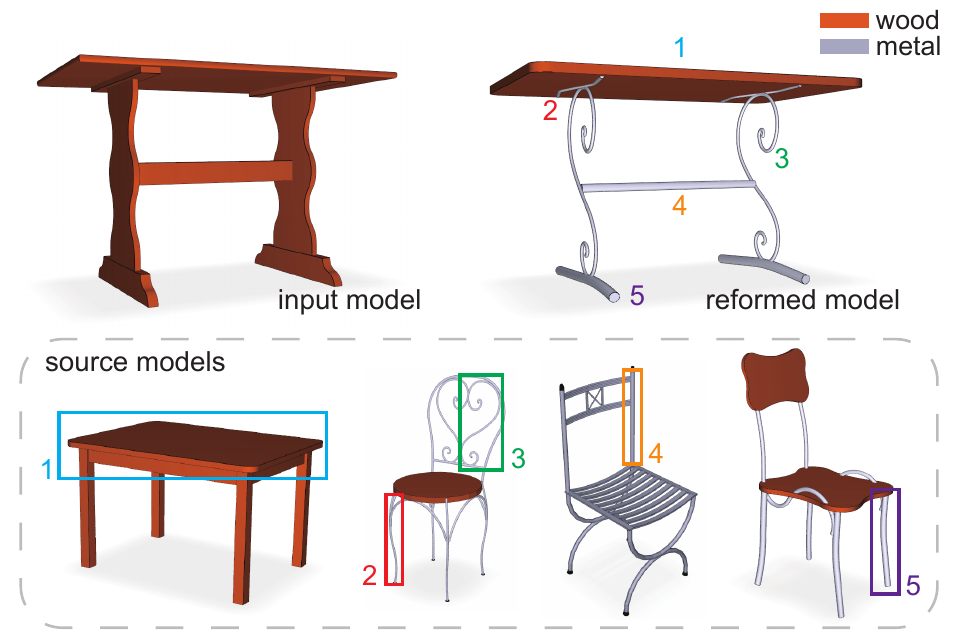}
  \caption{ A wooden table is reformed to have a metal structure based on example models in the database. The replacement parts are suggested using the proposed data-driven optimization.}
  \label{fig:reform}
  \vnudge
\end{figure}

\mypara{Formulation} For each part $P_n$ with target material $\bar{M}_n$, we define a potential $\phi(P_n^t)$ to measure the probability of replacing $P_n$ by $P_n^t$ (with material $M_n^t$):
\begin{equation}
\bar{\phi}(P_n^t) = \rho_{mat}(\bar{M}_n, M_n^t) \tilde{\rho}_{shp}(P_n, P_n^t),
\label{equ:reform_part}
\end{equation}
where $P_n^t \in \{P_1^t, P_2^t, ..., P_K^t\}$. Please note that here $\tilde{\rho}_{shp}(P_n, P_n^t)=\rho_{obb}(P_n, P_n^t)$, i.e., we only use OBB to measure the shape difference. Such relaxation allows suitable form changes to adapt to different materials. For example, a wooden board can be replaced by a curved metal part (see Fig.~\ref{fig:reform}). If part compactness is required (e.g., cabinets), we further add $\rho_{area}$ into $\tilde{\rho}_{shp}$.

For two parts $P_i$ and $P_j$ that are in contact, we define a pairwise potential $\bar{\psi}_{c}(P_i^t, P_j^t)$ to measure the probability of replacing $P_i$ by $P_i^t$, and $P_j$ by $P_j^t$. To do this, we compare with all the contacting pair $(P_u^t, P_v^t)$ in the training data set:
\begin{align}
\bar{\psi}_{c}(P_i^t, P_j^t)=\sum_{(u,v)} & \rho_{mat}(\bar{M}_i, M_u^t) \rho_{mat}(\bar{M}_j, M_v^t) \nonumber \\
          & \tilde{\rho}_{shp}(P_i^t, P_u^t) \tilde{\rho}_{shp}(P_j^t, P_v^t).
\label{equ:reform_contact}
\end{align}

For two congruent parts $P_i$ and $P_j$, we define a pairwise  potential $\psi_{r}(P_i^t, P_j^t)$ to encourage the same replacement part:
\begin{equation}
\bar{\psi}_{r}(P_i^t, P_j^t) = \delta_{ij}(P_i^t, P_j^t),
\label{equ:reform_congurent}
\end{equation}
where $\delta_{ij}(P_i^t, P_j^t)=1$ if and only if $P_i^t = P_j^t$, otherwise 0.

The overall potential of all part replacements $\set{P}^t$ is defined as:
\begin{equation}
F(\set{P}^t) = \prod_{P_n^t \in \set{P}^t} \bar{\phi}(P_n^t) \prod_{(i,j) \in \set{E}_c} \bar{\psi}_{c}(P_i^t, P_j^t)^{\alpha} \prod_{(i,j) \in \set{E}_r} \bar{\psi}_{r}(P_i^t, P_j^t)^{\beta},
\label{equ:reform_all}
\end{equation}
where $\alpha=0.1, \beta=20$ are weighting parameters. We optimize the above multi-label assignment problem by loopy belief propagation. The implementation is based on the sum-product algorithm~\cite{Kschischang:2001:FGS}.
Fig.~\ref{fig:reform_comparison} shows the effect of adding pairwise term.

Please note that in general, one can randomly specify metal or wood for each part, which often leads to unrealistic results. A better solution is to perform material suggestion based on example models using a similar approach as in~\cite{Jain:2012:MMA} (see Appendix~\ref{sec:material_assign}), and adapt model parts to different materials (i.e., metal$\rightarrow$wood, wood$\rightarrow$metal, see Fig.~\ref{fig:reform}).


%
%
%

\mypara{Restore contact}

\begin{figure}[t!]
  \centering
  \includegraphics[width=\columnwidth]{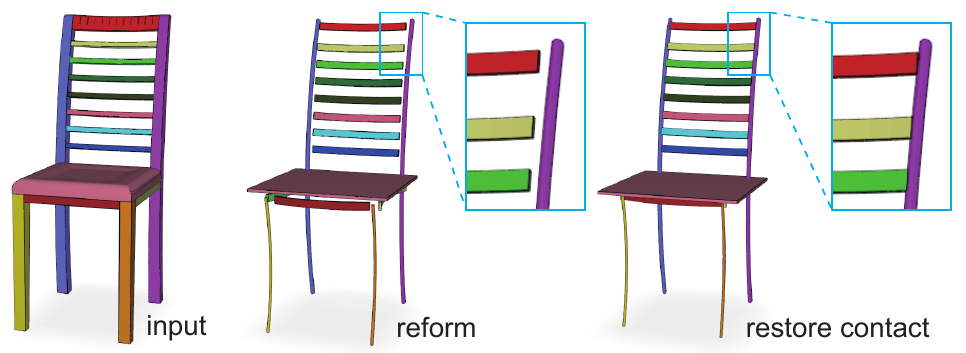}
  \caption{We perform a contact optimization to jointly connect replacement parts.}
  \label{fig:contact_opt}
  \vnudge
\end{figure}

After optimization, we find for each part $P_n$ a replacement part $P_n^r$. Now we want to use the query model $P=\{P_1, P_2, ..., P_N\}$ as reference to refine the position, orientation and scale of $\{P_1^r, P_2^r, ..., P_N^r\}$, so that they form a new model with corresponding parts in contact (see Fig.~\ref{fig:contact_opt}). First, we align $P_n^r$ to $P_n$ by aligning their OBB's. Then we scale $P_n^r$ so that the dominant dimension of its OBB matches $P_n$. Note that we only scale along the dominant dimension so that the thickness of $P_n^t$ is not affected. Then we fix the part with the most contacts, say $P_n$, and perform a joint optimization to optimize the location of all other parts to restore the contact graph of the original model. The basic idea is to use the original contact points to drive the replacement parts. In the contact graph, we have computed the contact points $\vec{c}_{ij}$ of neighboring parts $P_i$ and $P_j$. Now we project $\vec{c}_{ij}$ to the replacement part $P_i^r$ and $P_j^r$, and get two foot points $\vec{p}_{ij}^i$ and $\vec{p}_{ij}^j$. Suppose $\vec{t}_i$ is the displacement vector of $P_i^r$, we perform the following least squares optimization:
%
\begin{equation}
\min  \sum_{e_{ij}\in \set{E}_c} [(\vec{p}_{ij}^i+\vec{t}_i)-(\vec{p}_{ij}^j+\vec{t}_j)]^2, \quad s.t. \quad \vec{t}_n = 0.
\label{equ:contact}
\end{equation}
The closed form solution gives the displacement vector of each part (see Fig.~\ref{fig:contact_opt}).

\subsection{Optimizing for Material-aware Part Configurations}
Although we have considered pair-wise contact similarity in the material-aware shape reform stage, the spatial relation of neighboring parts still largely depends on the initial configuration. However, as mentioned before, how the parts should be manufactured and assembled is restricted by the fabrication material. Therefore, after reform, the updated model may have non-optimized structures which are not suitable for fabrication using the target material  . Here we propose an example based optimization which can further improve the fabrication feasibility of the reformed model.

\begin{figure}
\vnudge
  \centering
  \includegraphics[width=.95\columnwidth]{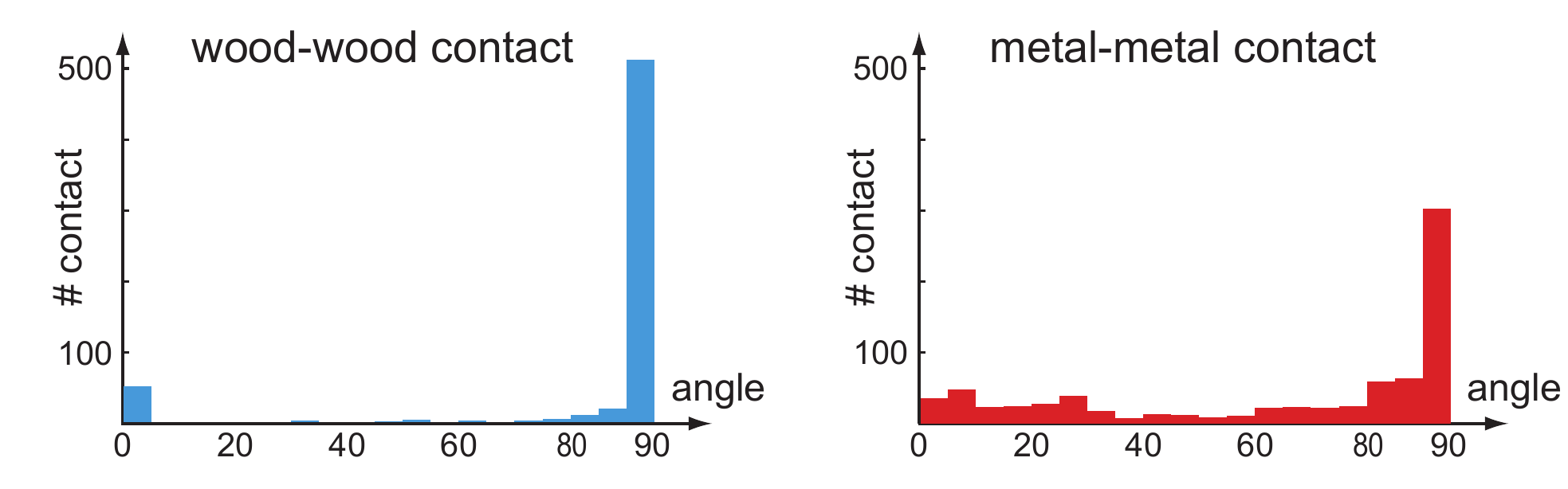}
  \caption{We build up histograms of contact angles for wooden parts and metal parts respectively, to infer fabrication feasibility and guide the structure optimization.}
  \label{fig:angle_hist}
\end{figure}

We use contact angle (see Sec.~\ref{subsec:similarity_metrics}) to measure the fabrication feasibility. From the examples in the database, we build up histograms of contact angles between wooden parts and between metal parts respectively (see Fig.~\ref{fig:angle_hist}). This captures the possibility/difficulty of constructing different contact angles in practice. Then from the reformed model, we compute contact angle between parts and measure its feasibility from the histogram. If the feasibility is low, we mark the contact and specify its target angle from the nearby feasible angles in the histogram. Given all the infeasible contacts denoted by edge set $\set{E}_{angle} \subset \set{E}_c$ in the contact graph, each infeasible contact $e_{ij} \in \set{E}_{angle}$ has two incident parts $P_i$ and $P_j$, a contact angle $\theta_{ij}$, and a target angle $\theta_{ij}^*$. We would like to perform an optimization to relocate the contacting parts so that all contact angles become feasible:
\begin{equation}
\min  \sum_{e_{ij} \in \set{E}_{angle}} \delta_{ij} \cdot [\theta_{ij}-\theta_{ij}^*]^2,
\;\;
\text{s.t.} \;\;   \sum_i \delta_{ij} \geq 2 \quad (\vec{c}_{ik} \neq \vec{c}_{il}).
\label{equ:angle_formulation}
\end{equation}
%
%
In the above equation, $\delta_{ij}$ is a binary variable. $\delta_{ij}=1$ indicates $P_i$ and $P_j$ are in contact, otherwise 0; $\vec{c}_{i,j}$ is the contact point of $P_i$ and $P_j$. The constraints ensure that there is no hanging part. We add an extra contact point for part who touches the ground (suppose the up-right orientation of the model is along $z$-axis).

\begin{figure}[b!]
  \centering
  \includegraphics[width=.95\columnwidth]{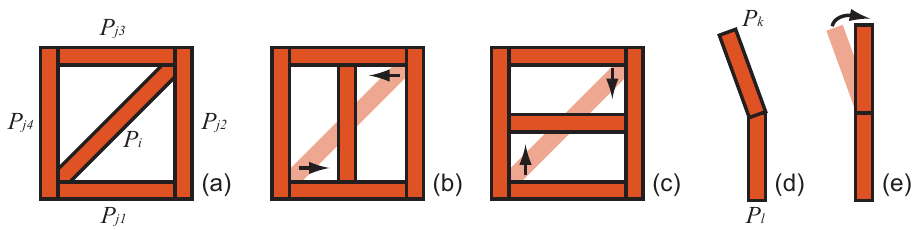}
  \caption{To optimize the contact feasibility, we allow part to slide/rotate on other parts. (a) $P_i$ cannot be orthogonal to all other four parts; (b), (c) by relaxing contact and allowing $P_i$ to slide, two feasible configurations can be achieved; (d) $P_k$ is allowed to rotate while fixing the contact point with $P_l$; (e) feasible configuration after rotation. }
  \label{fig:angle_illust}
  \vnudge
\end{figure}

Ideally, we should have $\delta_{ij}=1, \forall e_{ij} \in \set{E}_{angle}$ in the optimization. However, in practice it is not feasible to converge to all target angles. One simple case is shown in Fig~\ref{fig:angle_illust}. This is mainly because metal allows much more contact freedom than wood. Instead, we relax the contacts by allowing $\delta_{ij}=0$. On the other hand, to keep contact between $P_i$ and $P_j$, we cannot just set $\delta_{ij}=1$, since $\delta_{ij}$ depends on the locations of $P_i$ and $P_j$.
Instead, we ensure contact by adding geometric constraints that only allow $P_i$ to slide on $P_j$ (i.e, changing contact point), or rotate while keeping contact point with $P_j$. Whether a part should slide or rotate depends on its relation with neighboring part (e.g., contact point location, target angle).

To simplify the computation, we abstract each (elongated) part $P_i$ by a line segment $s_i=(\vec{v}_i^s, \vec{v}_i^t)$ (using two end points along dominant dimension of its OBB). Given all the contact constraints, i.e., sliding pairs $ \{e_{ij}\} \subset \set{E}_{angle}$ ($P_i$ slides on $P_j$), rotating pairs $ \{e_{kl}\} \subset \set{E}_{angle}$ ($P_k$ rotates while contacting $P_l$), the formulation in Eqn.~\ref{equ:angle_formulation} can be interpreted as:
\begin{align}
\min & \sum_{e_{ij}} [\theta( s_i, s_j )-\theta_{ij}^*]^2
+ \sum_{e_{kl}} [\theta( s_k, s_l )-\theta_{kl}^*]^2
\nonumber \\
& + w_{l}\sum_{e_{kl}}[(\vec{v}_k^s-\vec{v}_k^t)^2 - (\hat{\vec{v}}_k^s-\hat{\vec{v}}_k^t)^2]^2, \nonumber \\
& + w_{r}\sum_{(u, v)} \sum_{(m, n)} Repulse( s_m, s_n ), \nonumber \\
& e_{mu}, e_{mv}, e_{nu}, e_{nv} \in \{e_{ij}\} \nonumber \\
\text{s.t. } \, & \vec{v}_i^* = t_{ij}\vec{v}_j^s + (1-t_{ij})\vec{v}_j^t, \nonumber \\
& 0 \leq t_{ij} \leq 1,  \quad \quad
 \vec{v}_k^* = \vec{c}_{kl},
\label{equ:angle_opt}
\end{align}
where $*\in\{s, t\}$ indicating which end of the part should slide/rotate, can be inferred from the initial configuration. We also add length preserving term for rotating parts. If $P_m$ and $P_n$ both slide on $P_u$ and $P_v$, a repulsion term $Repulse( s_m, s_n )=\exp[-(\vec{v}_m^s - \vec{v}_n^s)^2 / \sigma^2] \cdot \exp[-(\vec{v}_m^t - \vec{v}_n^t)^2 / \sigma^2]$ is added to avoid part overlap.

In our implementation, we enumerate all the slide/rotate possibilities, specify the corresponding constraints, and perform numerical optimization to solve Eqn.~\ref{equ:angle_opt} (with $w_l=1, w_r=0.1, \sigma=0.05$). The solution which reaches minimal objective function value is selected as the optimal result.
The model is updated accordingly by aligning parts to the corresponding line segments.

\begin{figure}[h]
  \centering
  \includegraphics[width=\columnwidth]{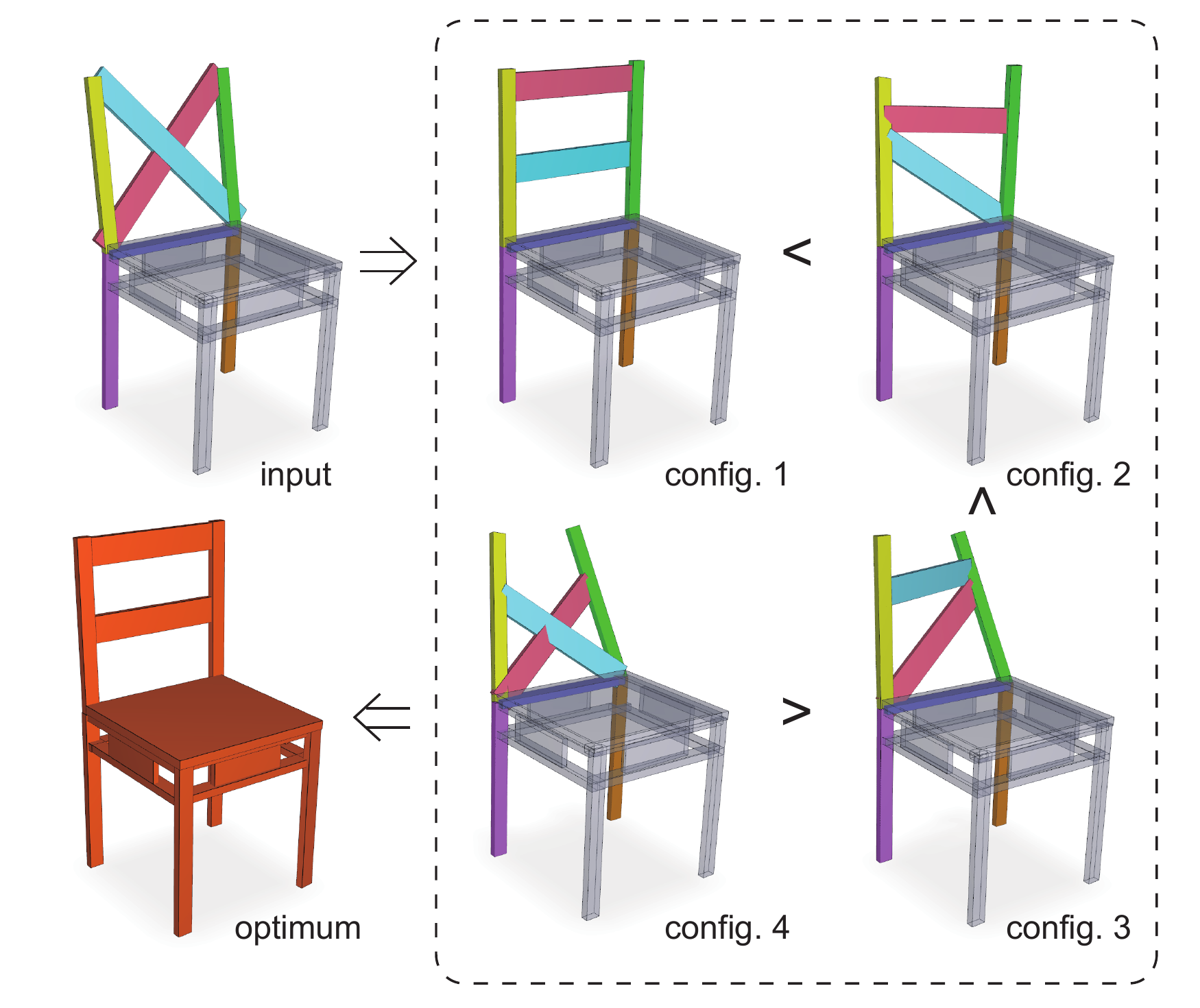}
  \caption{Material-specific optimization on a reformed chair model. We use contact angle histogram to detect non-optimal contact angles. The incident parts are shown in dark colors. We optimize the angle differences between parts under different contact constraints, resulting in different configurations. We sort the configurations based on contact angles and select the optimum one. }
  \label{fig:angle_sampling}
  \vnudge
\end{figure}

To initialize the optimization, we fix the parts that have no contact part at one of its end, or have symmetric parts with no angle problem. We also restrict two ends of a part to slide on non-contacting parts. This allows the optimization to run faster, and explore more interesting variations with topology and geometry change (see Fig.~\ref{fig:angle_sampling}).

\subsection{Generating Fabrication Specifications}
\begin{figure}
\vnudge
  \centering
  \includegraphics[width=\columnwidth]{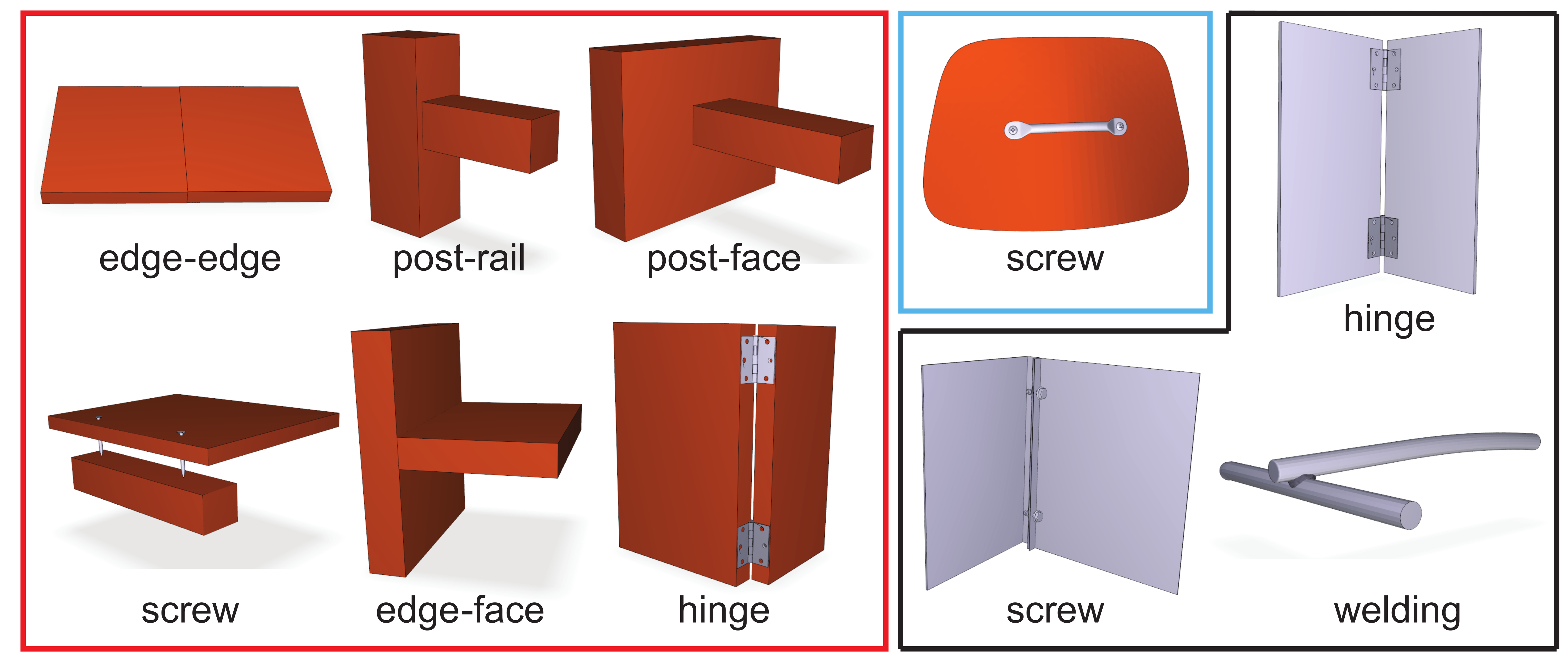}
  \caption{Representative joint types of wood-wood (red box), wood-metal (blue box) and metal-metal (black box). }
  \label{fig:joints}
\end{figure}

So far we have generated the geometry of individual parts and optimized their configurations according to target materials. In practice, how to assemble multiple parts to form a real functional object is a non-trivial task~\cite{Hylton08}. It not only depends on the materials of neighboring parts (e.g., metal parts are connected by welding, wooden parts are connected by specific wooden joints), but also their geometry and spatial correlation (see Fig.~\ref{fig:joints}). In this section, we would like to further investigate the fabrication specifications in terms of how neighboring parts are conjoined. The goal is to infer joint type based on part geometry, and further optimize/refine its geometry (for wooden parts) to ensure the overall assembly.

\mypara{Fabrication type inference}
The actual fabrication of a multi-component object requires different conjoining methods to assemble neighboring parts. Fig.~\ref{fig:joints} shows several representative joint types that are resolved in our framework. These joint types are classified into three categories (i.e., wood-wood, wood-metal and metal-metal) based on the material configuration of the neighboring parts. For each category, different fabrication techniques can further be applied to assemble parts with different characteristics.

Similar to shape reform, we also perform a data-driven approach to infer different joint types for the reformed model. Namely, we derive joint type by comparing with example models with pre-tagged joints. Suppose $P_i$ and $P_j$ are the incident parts of the query joint, the potential of assigning the same type of joint as $P_u^t$ and $P_v^t$ is expressed as:
\begin{align}
\phi_{i,j;u,v} = & \rho_{mat}(M_i, M_u^t) \rho_{mat}(M_j, M_v^t) \nonumber \\
          & \tilde{\rho}_{shp}(P_i, P_u^t) \tilde{\rho}_{shp}(P_j, P_v^t) \nonumber \\
          & \rho_{pr}(d_{i,j}, d_{u,v}) \rho_{oa}(\vec{a}_{i,j}, \vec{a}_{u,v}),
\end{align}
where $\vec{a}_{i,j}$ is a 9-d vector that measures the angles ($\in [0, 90]^{\circ}$) between individual principal directions of two OBB's. Specifically, suppose the principle directions of $P_i$ is $\{{X}_{i0},{X}_{i1},{X}_{i2}\}$ (in order), then $\vec{a}_{i,j}$ is the vector representation of the $3\times3$ matrix $\{Angle(X_{ik},X_{jl})\}_{k,l=0,1,2}$ and $\rho_{oa}(\vec{a}_{i,j}, \vec{a}_{u,v}) = \exp(\frac{-|\vec{a}_{i,j} - \vec{a}_{u,v}|^2}{\sigma_{oa}^2})$, where $\sigma_{oa} = 360^{\circ}$ in our experiment.
The definition of the other terms can be found in Sec.~\ref{subsec:similarity_metrics}.

\begin{figure}[b!]
  \centering
  \includegraphics[width=.95\columnwidth]{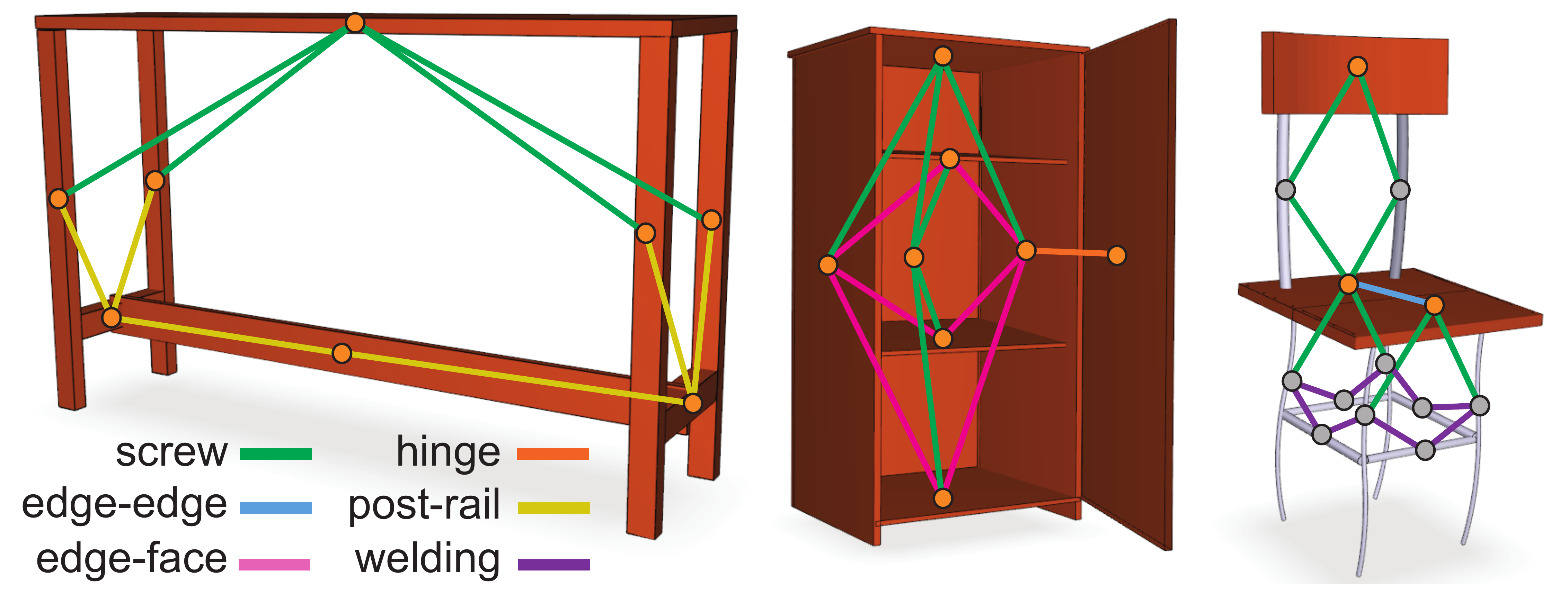}
  \caption{Different fabrication joint types (shown as graph edges) are inferred using a data-driven approach.}
  \label{fig:fab_infer}
  \vnudge
\end{figure}

\begin{figure}[h]
  \centering
  \includegraphics[width=\columnwidth]{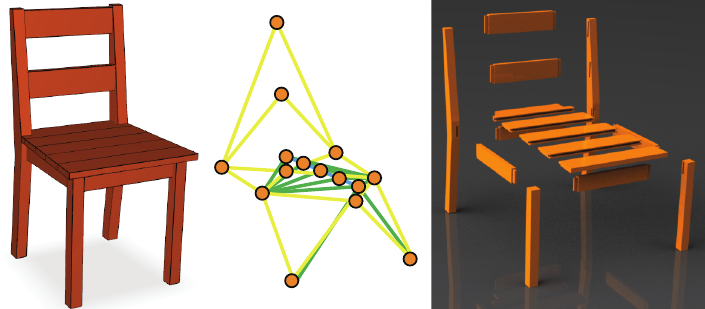}
  \caption{Fabrication joints can be created based on the right joint type and part geometry. Please refer to Figure~\ref{fig:angle_sampling} for edge-color legend. }
  \label{fig:fab_joints}
  \vnudge
\end{figure}

For joint assignment, we classify all pre-tagged joints into multiple clusters based on the specified joint types. The type of the query joint is inferred by its $k$ nearest neighbors among the pre-tagged models from different clusters (see Fig.~\ref{fig:fab_infer}). Please note that our joint assignment is entirely geometry-based. It doesn't rely on any semantic information of the models/parts (up-right/front orientation, part labels, etc.). For functional parts, this may lead to ambiguities. For example, a closed cabinet door can not be separated from cabinet back. Then the user is expected to clarify such ambiguities.
After the assignment, we further verify tenon/mortise part (tenon tongue or mortise hole will be generated accordingly, see Fig.~\ref{fig:fab_joints}) based on pairwise relations (e.g., the contact location) for later processing.

\vspace{-0.1in}
\mypara{Fabrication-aware part refinement}
Based on the inferred joint types, we further seek to create tangible joint
\begin{wrapfigure}[15]{r}{.31\columnwidth}
\vspace{-10pt}
\includegraphics[width=.3\columnwidth]{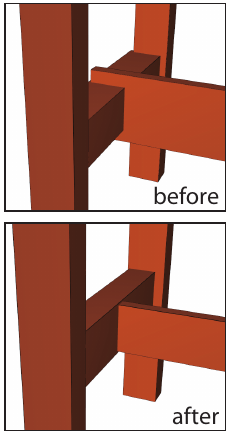}
\end{wrapfigure}
shapes by refining neighboring wooden parts. However, the dimensions of the neighboring parts may not be compatible after shape reform (see inset figure), making the actual fabrication unrealizable. We thus perform a proxy-based deformation~\cite{Zheng11} to resize neighboring parts and ensure the right configuration. We use part's OBB as proxy and solve for the optimal dimensions of the OBB (the centroid is kept fixed). We optimize the closeness to the original OBB's subject to compatibility constraints (e.g., the tenon part should lie in the mortise part along major axis) on neighboring OBB's. This results in a quadratic function (in terms of OBB dimensions) with linear constraints, which can be explicitly solved by Quadratic Programming. After optimization, all the related parts are updated (scaled) according to the OBB's (see inset figure).


\mypara{Forming joints}
With the right joint type and part geometry, we form joint shapes by sculpting corresponding wooden parts to ensure the assembly after fabrication. This is based on several simple CSG operations. First, we use mortise part to subtract tenon part and detect the contact face on the tenon part. Then we construct the tenon geometry by scaling the contact face and extending to the mortise part. Finally, the tenon/mortise part is sculpted by adding/subtracting the tenon geometry (see Fig.~\ref{fig:fab_joints}).

\begin{figure*}[t!]
  \centering
  \includegraphics[width=0.98\linewidth]{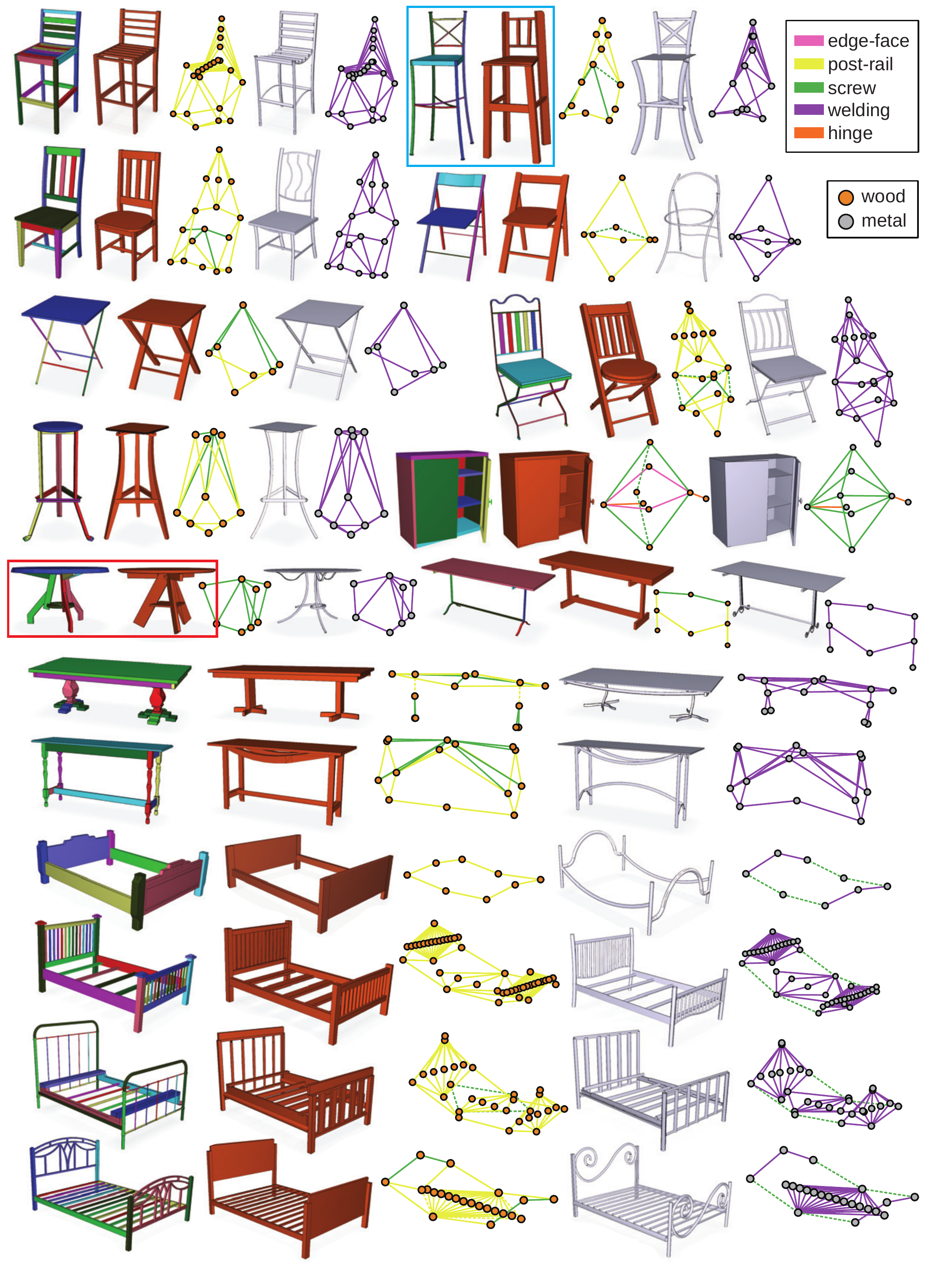}
  \caption{Fabrication-aware shape reform on a number of models. For each model, the target material is set to wood and metal respectively. The fabrication joint inference result is shown next to the reformed model. Ambiguous joints are shown by dash lines.}
  \label{fig:reformall}
\end{figure*}

\begin{figure}
  \centering
  \includegraphics[width=0.95\linewidth]{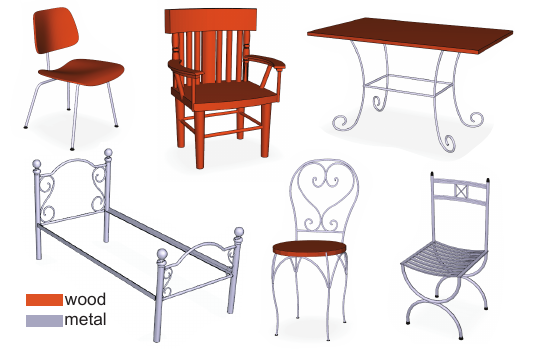}
  \caption{Fabrication materials are automatically suggested for each part of the model by learning from a database.}
  \label{fig:classificationall}
\end{figure}

\section{Evaluation}
In this section, we evaluate the proposed framework. First, we test our algorithm using different settings on a large number of query models. Then we explore several reform variations. Further, we show the conducted user study and the statistics of our framework. All results are generated with the whole database unless mentioned otherwise. The database models are mixed of chairs (83 models, 1304 parts), tables (32 models, 350 parts), beds (19 models, 588 parts) and cabinets (18 models, 221 parts).



\begin{figure}[b!]
  \centering
  \includegraphics[width=0.94\linewidth]{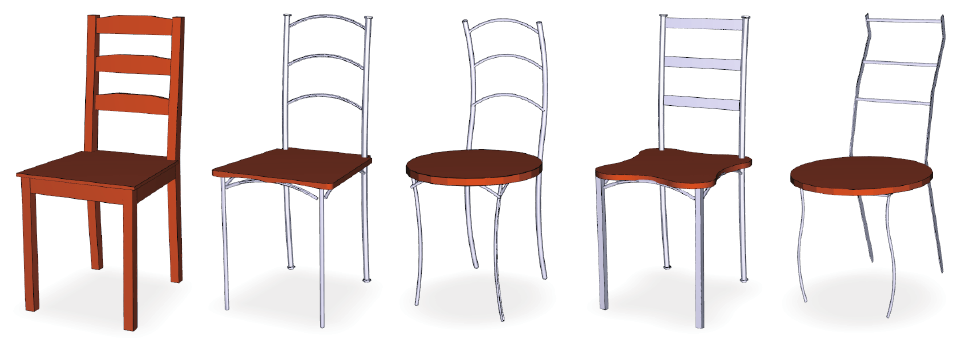}
  \caption{Given an input model (on the left), multiple reform results can be generated from different example dataset.}
  \label{fig:subset}
  \vnudge
\end{figure}

\mypara{Results}
We test our material-aware reform algorithm (coupled with material specific optimization) on a number of models (see Fig.~\ref{fig:reformall}). To show the generality of the proposed method, here we reform the input model to have only wooden/metal parts.

We achieve plausible reformed models which successfully adapt to the target material constraints (see also user study and video/demo). Interesting reform patterns can be well observed. For example, in `to metal' cases, straight bar become curved arc, flat boards become planar snakes, the thickness of the reformed parts is small (opposite effects can be observed in `to wood' case). All these patterns conform to the correlations between built material and geometric forms (c.f., Fig.~\ref{fig:fabrication_angle}). If the target material happens to be the underlying material of the input model (we do not enforce reform to different material here), the resultant model usually has a comparable structure with the input model, while exhibiting feasible geometric variations. More importantly, as marked by the blue box (see also Fig.~\ref{fig:teaser}, Fig.~\ref{fig:overview}), our framework can also optimize the inter-part topology of the reformed model, to facilitate fabrication in the target material context. For example, part connections are broken at the chair back and leg, so that wooden parts can be optimized to meet at near $90^{\circ}$. On the other hand, we also find some challenging case as shown in red box. This is not only caused by inappropriate modeling/segmentation of the input table legs, but also the complicated spatial configurations of the output. A small amount of user interactions may further be required for such cases.

Besides replacing and optimizing individual parts, fabrication joints on the reformed models are also inferred from the database (see Fig.~\ref{fig:reformall}). Our data-driven approach can resolve most of the joint types. Ambiguous joints are mainly caused by the lack of shape semantics. For example, for metal bed, bed top and bed body need to be screwed to ease assembly/disassembly.


\mypara{User control}
As mentioned before, one option for shape reform is to use material suggestion (see Appendix~\ref{sec:material_assign}) to infer the current material context, and adapt model parts to different materials. Fig.~\ref{fig:classificationall} shows automatic material suggestion for various query models. Statistical numbers are discussed later. The rest reform results in the paper, including mixed target materials, are generated referring to the current material context.

Given a single input model and the target material of each part, material-specific reform variations can be achieved by exploring different subspaces in the database. Fig.~\ref{fig:subset} shows several reformed metal chairs from the same wooden chair. Each reformed model is generated by randomly selecting 20 chairs in the database as a new example dataset.

\begin{figure}[t!]
  \centering
  \includegraphics[width=0.99\linewidth]{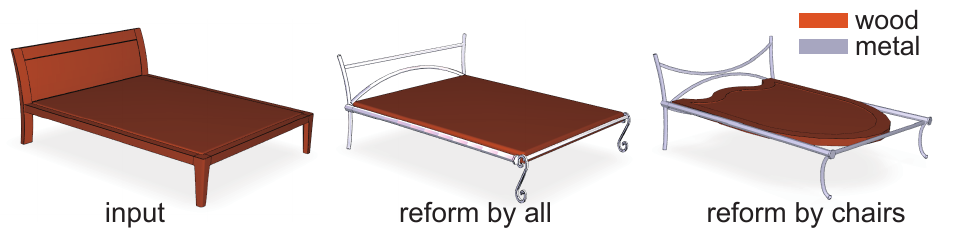}
  \caption{Given a bed with all wooden parts (left), we can reform for mixed materials, e.g., the bed frame adapts to metal and the bed board adapts to wood. The reformed result from the whole database (middle). The reformed result from only the chair dataset (right).}
  \label{fig:change_database}
  \vnudge
\end{figure}

\begin{figure}[b!]
  \centering
  \includegraphics[width=0.90\linewidth]{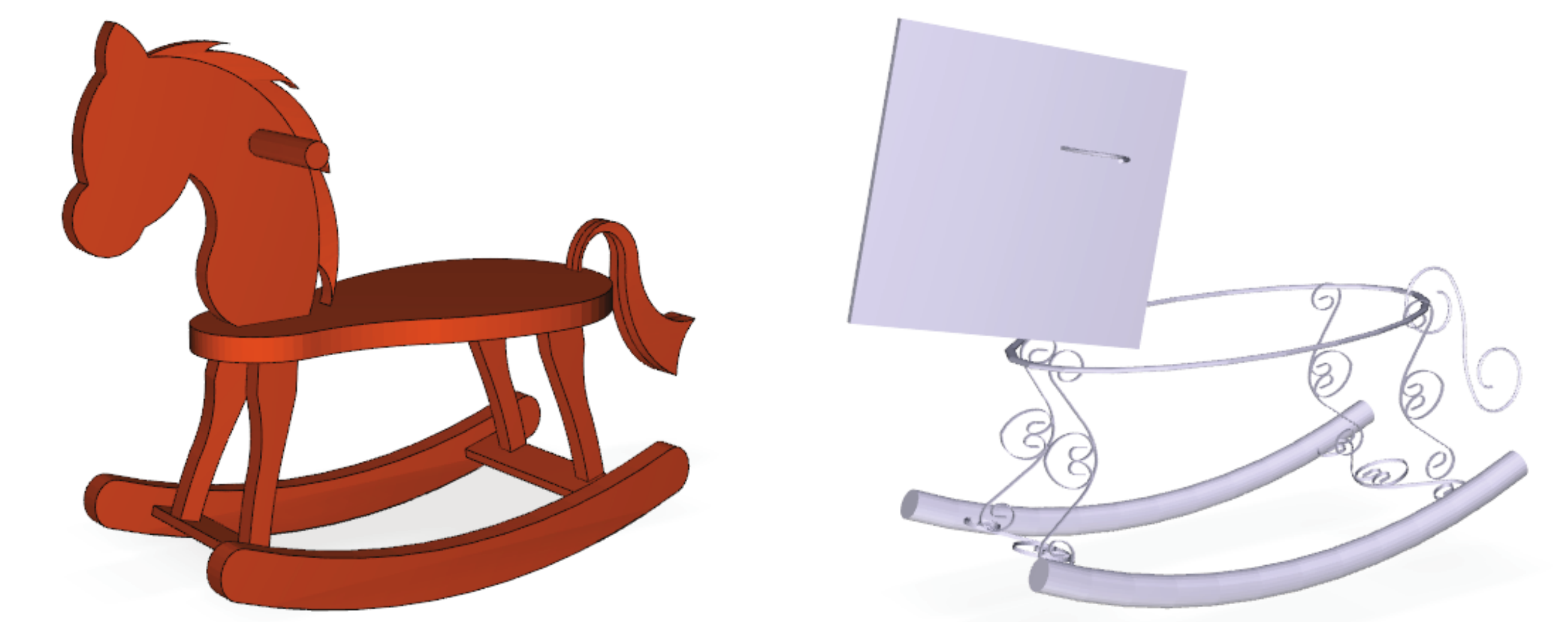}
  \caption{Plausible result generated by reforming a toy horse.}
  \label{fig:toyhorse}
  \vnudge
\end{figure}

Fig.~\ref{fig:change_database} shows a shape reform result across different type of models in the database. It is easy to see that the bed board cannot find a compatible part if only search in a chair database. Stretching a chair seat to be a bed board results in large distortion. On the other hand, forms with the same material share common features across different type of models. For example, compatible metal parts can be retrieved from a chair dataset to form a plausible metal bed frame.
Fig.~\ref{fig:toyhorse} shows a shape reform result from a toy horse.


\mypara{User study}
We conducted a user study to assess the results of our framework. We showed 35 material-aware shape reform results (the test models are randomly picked from our database, see supplemental material) to 30 computer science and EE students with varied background. For each reformed model, the users were asked to infer its major built material from the geometry alone (we did not  show the material color to the user).

\begin{figure}[t!]
  \centering
  \includegraphics[width=0.94\linewidth]{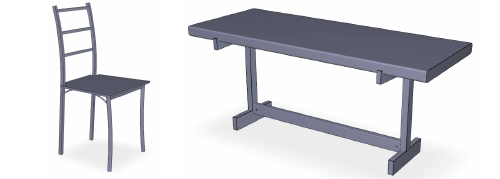}
  \caption{It is not easy to recognize material from geometric shape alone if material characteristics cannot be inferred from individual parts and part configurations (e.g., contact angles). Left: The metal parts of the model are mainly straight and most of the contact angles are 90 degrees. It accords with typical wooden chairs and makes people believe it is made of wood. Right: The wooden parts of the table are thick, while the whole structure is still similar to a typical metal table. }
  \label{fig:user_study}
  \vnudge
\end{figure}

The feedback was largely positive. The average hit rate was 91.5\%. The reformed shape thus conformed to human expectation in material and geometric form.
Fig.~\ref{fig:user_study} shows two typical failure cases where users' choice did not match the specific material.

\begin{figure}[h]
  \centering
  \includegraphics[width=\linewidth]{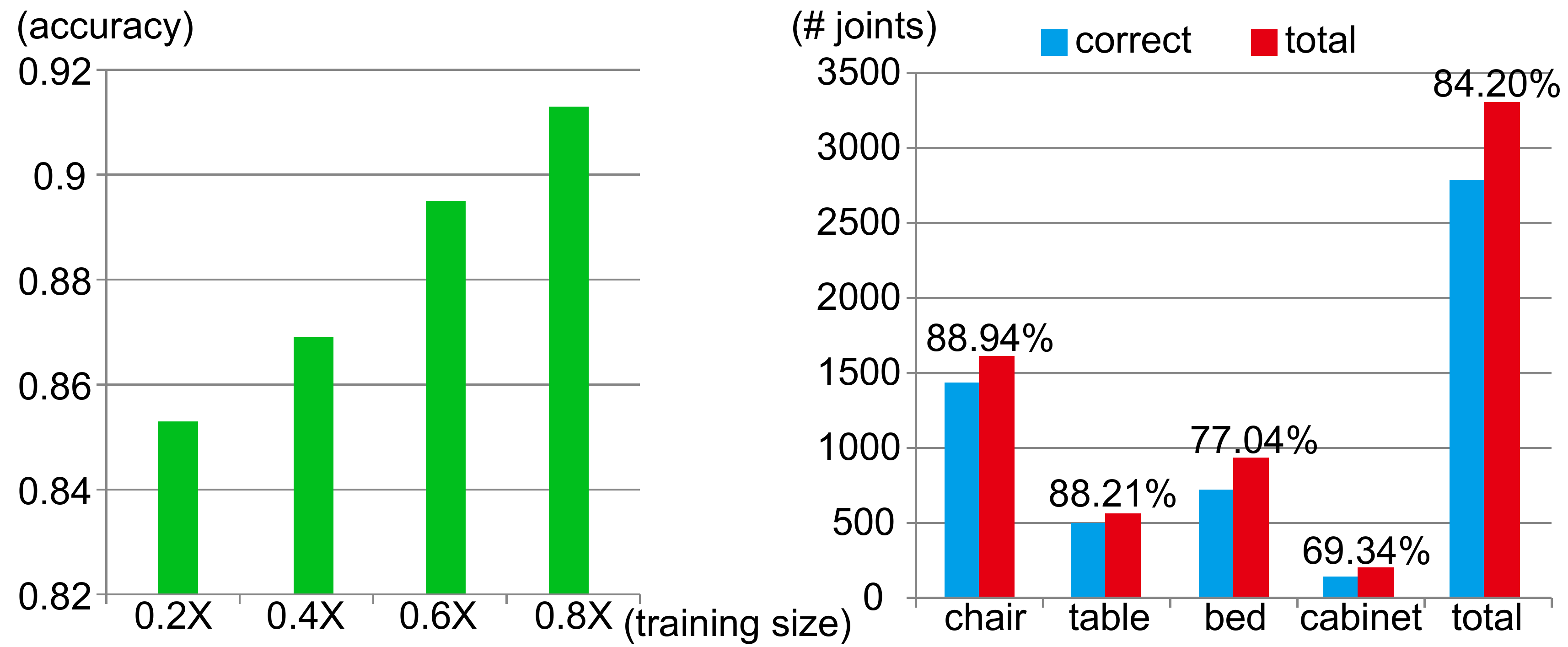}
  \caption{Left: the accuracy of predicting fabrication materials using different proportions of the data base as training set. Right: the accuracy statistics of predicting fabrication joint types.}
  \label{fig:statistics}
  \vnudge
\end{figure}

\vspace{-0.1in}
\mypara{Statistics}
Cross validation of the material suggestion is tested on different training data sets. We randomly sample different subsets (with ratio 0.2, 0.4, 0.6, 0.8) from the database as training data and use the rest as testing data. For each ratio, we run 10 times and select the top 4 material suggestion accuracy, then compute the average. The statistics is shown in Fig.~\ref{fig:statistics} (left). We also perform fabrication type inference on all the models in the database. The query model itself is leaving out in the training set during the inference process. The inference accuracy numbers of individual categories are summarized in Fig.~\ref{fig:statistics} (right).

\vspace{-0.1in}
\mypara{Performance}
Our experimental platform is with a 2.66 GHz Intel Xeon X5550 CPU. The training of the database is computed offline. Since we only have two candidate materials (metal/wood), it only takes 4.6 seconds to assign materials for a model with 27 parts. For material-aware reform, if we consider all parts in the database to be candidates, the optimization is very expensive. We handle this problem in a preprocessing step. First, we filter out congruent parts (keeping only one instance). Then we cluster the rest parts into 80 groups by $k$-means clustering using feature vector of each part as its OBB dimension, area ratio and thickness (5D in total). For each cluster, the part with the smallest distance to the cluster center is selected as a candidate part. The reform based on clustered candidate parts can be computed efficiently, it takes 15.3s to reform the same model with 27 parts. The fabrication inference takes 3.9s for 55 contacts of the same model.

\section{Conclusion}

We presented a data-driven algorithm that reforms a component-based input shape such that the reformed shape is better suited for fabrication using the target build material. We formulated this as an optimization that not only selects appropriate parts from the database, scales and positions them appropriately, but searches over non-trivial topological changes to ensure that the reformed shape conform to material-specific angle distributions. Finally, part connection types are inferred and necessary geometric modifications are suggested. We validated the algorithm on various models for wood and metal constructions, and evaluated the results using a user study that the  classification results and reformed shapes match human perception of material just based on geometric shape.

{\em Limitations and future work:}
Our reform algorithm is based on a part replacement strategy. Since different materials result in different part shaping abilities, if one part cannot
\begin{wrapfigure}[16]{r}{.22\columnwidth}
\hspace{-8pt}
\vspace{-14pt}
\includegraphics[width=.22\columnwidth]{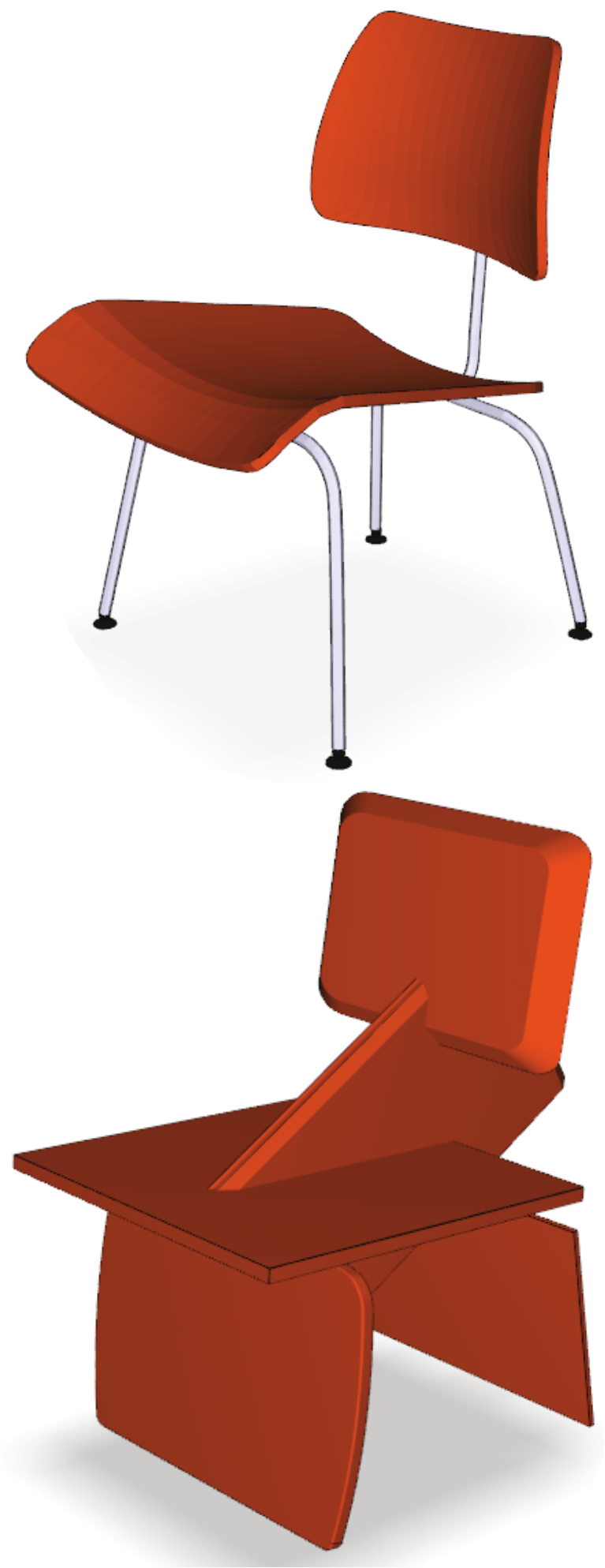}
\vspace{-30pt}
\end{wrapfigure}
 find a geometrically similar part when adapting to a different target material, unplausible result would be generated (see inset figure).
Also, in this work we only focused on wood and metal materials. In the future, it would be interesting to extend the
framework to also handle plastic and molded sheets. However, the challenge then would be to obtain initial
parts since the input geometric meshes do not necessarily conform to material specific partitioning.
Our algorithm does not assume access to part labels (table leg, chair seat, etc.). As a result
parts can potentially undergo large deformations, say a chair leg can get stretched to become a bar for the bed frame.
In reality, however, materials have limits on maximum dimensions. It is desirable for the algorithm to take this into account when making part suggestions.
Finally, we did not assume models to contain part connector geometry (e.g., nails, bolts, rivets). When available, it will be beneficial to infer them and redirect our classifications.

\appendix

\section{Material Suggestion (optional)}
\label{sec:material_assign}


Given a model $\mathcal{P}$ consists of individual parts $\{P_1, P_2, ... , P_N\}$, in an optional initialization step, fabrication material $M_n \in \{$metal, wood$\}$ is assigned for each part $P_n, (1 \leq n \leq N)$. This helps to adapt model parts to different materials in the shape reform stage.
\mypara{Formulation}
For each part $P_n$, we define a potential $\phi(M_n)$ to measure the probability of assigning material $M_n$, by comparing with all the parts in the training set $\{P_1^t, P_2^t, ..., P_K^t\}$ with materials $\{M_1^t, M_2^t, ..., M_K^t\}$:
\begin{equation}
\phi(M_n)=\sum_{k=1}^{K} \rho_{mat}(M_n, M_k^t) \rho_{shp}(P_n, P_k^t),
\label{equ:mat_part}
\end{equation}
where $\rho_{shp}(P_n, P_k^t)$ is a combination of $\rho_{obb}$, $\rho_{area}$ and $\rho_{thick}$.

For two parts $P_i$ and $P_j$ that are in contact, we define a pairwise potential $\psi_{c}(M_i, M_j)$ to measure the probability of assigning material $M_i$ to $P_i$, and $M_j$ to $P_j$, by comparing with all the contacting pairs $(P_u^t, P_v^t)$ in the training data set:
\begin{align}
\psi_{c}(M_i, M_j)=\sum_{(u,v)} & \rho_{mat}(M_i, M_u^t) \rho_{mat}(M_j, M_v^t) \nonumber \\
          & \rho_{shp}(P_i, P_u^t) \rho_{shp}(P_j, P_v^t) \nonumber \\
          & \rho_{pr}(d_{i,j}, d_{u,v}) \rho_{ca}(\alpha_{i,j}, \alpha_{u,v}).
\label{equ:mat_contact}
\end{align}

For two congruent parts $P_i$ and $P_j$, we define a pairwise potential $\psi_{r}(M_i, M_j)$ to encourage the same material suggestion:
\begin{equation}
\psi_{r}(M_i, M_j) = \rho_{mat}(M_i, M_j).
\label{equ:mat_congruent}
\end{equation}

\begin{figure}[t!]
  \centering
  \includegraphics[width=.95\columnwidth]{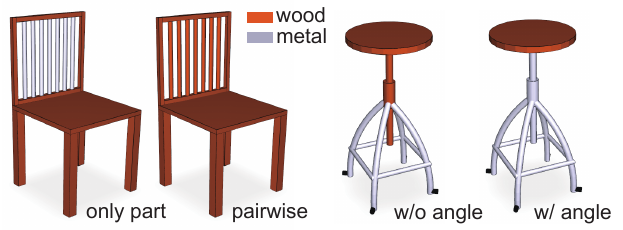}
  \caption{(Left)~Pairwise factor, which is based on comparing neighboring configuration in a database, helps to suggest right material configuration between parts.
  (Right)~Additionally, inter-part angle similarity can help to make more appropriate material assignment to parts that are easier to fabricate.}
  \label{fig:material_assignment}
\end{figure}

For the whole model, the potential of the material suggestion $M$ is defined as:
\begin{equation}
F(M) = \prod_{M_n \in M}\phi(M_n) \prod_{(i,j) \in \set{E}_c}\psi_{c}(M_i, M_j)^{\alpha} \prod_{(i,j) \in \set{E}_r}\psi_{r}(M_i, M_j)^{\beta},
\label{equ:mat_all}
\end{equation}
where $\alpha=0.1, \beta=20$ are weighting parameters. The optimal material suggestion is solved by loopy belief propagation. Fig.~\ref{fig:material_assignment} shows the effect of pairwise factor and angle similarity.


\bibliographystyle{acmsiggraph}
{\fontsize{8.1pt}{8.8pt}\selectfont
\bibliography{geomat}
\end{document}